УДК 316.776:004.58

**Соколов Володимир Юрійович**
старший викладач кафедри інформаційної та кібернетичної безпеки
Київський університет імені Бориса Грінченка, Київ, Україна
OrcID: 0000-0002-9349-7946
*vladimir.y.sokolov@gmail.com*

**Курбанмурадов Давид Миколайович**
магістр
Державний університет телекомунікацій, Київ, Україна
OrcID: 0000-0003-4503-3773
*rockyou@protonmail.com*

## МЕТОДИКА ПРОТИДІЇ СОЦІАЛЬНОМУ ІНЖИНІРИНГУ НА ОБ'ЄКТАХ ІНФОРМАЦІЙНОЇ ДІЯЛЬНОСТІ

**Анотація.** В статті приведено дослідження з використанням атак типу підробленої точки доступу та фішингової сторінки. Оглянуті попередні публікації з соціального інжинірингу (соціальної інженерії), приведена статистика зламів і проаналізовані напрямки і механізм реалізації атак, що мають елементи соціального інжинірингу. Зібрано та проаналізовано дані з дослідження у трьох різних місцях та наведено змістовну статистику. Для порівняння були вибрані три категорії вищих навчальних закладів: технічного, гуманітарного і змішаного профілів. Так як дослідження проводилися у навчальних закладах протягом тижня, то в більшості своїй в експерименті взяли участь студенти й аспіранти. Для кожного навчального закладу був зроблений свій шаблон форми реєстрації, який імітував дизайн головних сторінок. Наведені приклади апаратно-програмної реалізації типового стенду для проведення атаки, збору та аналізу даних. Для побудови дослідного стенду були вибрані широкодоступні компоненти, щоб показати, наскільки легко проводити атаки подібного роду без великих початкових витрат і спеціальних навичок. В статті приведена статистика з кількості підключень, дозволу на використання адреси електронної пошти і паролю, а також дозволу на автоматичне передавання службових даних браузером (кукі). Статистичні дані оброблені за допомогою спеціально написаних алгоритмів. Запропоновані підходи до вирішення проблеми соціотехнічних атак можуть бути використані та впроваджені для експлуатації на будь-яких об'єктах інформаційної діяльності. В результаті експериментів видно, що обізнаність користувачів навіть технічних спеціальностей недостатня, тому потрібно приділяти окрему увагу до розробки методик підвищення рівня обізнаності користувачів та зменшення кількості потенційних атак на об'єкти інформаційної діяльності.

**Ключові слова:** соціальний інжиніринг; соціальна інженерія; атака; фішинг; точка доступу; захист персональних даних.

## 1. ВСТУП

В наш час усі підприємства пов'язані з процесами зберігання та обробки інформації. Ця інформація може містити конфіденційні дані, розкриття яких завдасть значної шкоди репутації підприємства, його роботоспроможності або фінансового положення.

Соціальний інжиніринг спрямований не на комп'ютерну техніку, а на користувача. Інтерес представляють всі платоспроможні особи, а також користувачі, що володіють цінною інформацією, співробітники підприємств і державних установ.

Цей метод застосовується з метою виконання фінансових операцій, зламу, крадіжки даних, наприклад, клієнтських баз, персональних даних і іншого несанкціонованого доступу до інформації. Соціальний інжиніринг допомагає конкурентам здійснювати розвідку, виявляти слабкі сторони організації, переманювати співробітників.

Питання соціального інжинірингу у вітчизняних наукових колах з'явився в часи різкого збільшення доступності інформаційних ресурсів, телекомунікаційних мереж і терміналів користувачів. Принципи впливу на людину за допомогою соціального інжинірингу приведені в [1]. Загальні питання витоку даних розглянуті в [2] і [3], безпосередньо соціального інжинірингу — у роботах [4] і [5]. Проблема надання доступу до службової інформації розглянута у [6].

Статистика зламів інформаційних систем по всьому світу станом на 2018 рік за даними компанії Verizon Communications Inc. представлена на рис. 1 [7]. За даними цієї статистики соціальний інжиніринг займає третє місце за кількістю атак.

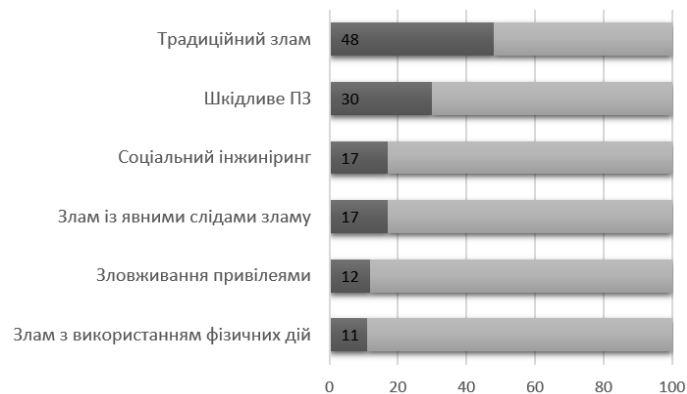

*Рис. 1. Статистика тактик зламу*

Існує багато джерел загроз інформаційній та кібербезпеці підприємства. До процесу зберігання та обробки інформації завжди залучений персонал підприємства. Отже, важливо роздивлятись антропогенний фактор як реально існуючу уразливість у інформаційній безпеці підприємства. За статистичними даними соціальний інжиніринг є найвагомішою загрозою, направленою на антропогенний фактор. Існує багато методів протидії методам соціального інжинірингу. Одним з таких методів є підвищення обізнаності персоналу у питаннях протидії методам соціального інжинірингу. Не всі підприємства приділяють належну увагу до підвищення обізнаності. Тому постає необхідним створення ефективної методики підвищення обізнаності персоналу у питаннях протидії методам соціального інжинірингу.

*Метою роботи* є підготовка обґрунтованої методики підвищення рівня обізнаності персоналу у питаннях протидії методам соціального інжинірингу. *Об'єктом дослідження* є процес управління обізнаністю персоналу. *Предметом дослідження* є протидія методам соціального інжинірингу.

*Наукова новизна* роботи полягає у розробці методики управління обізнаністю персоналу у питаннях протидії методам соціального інжинірингу. *Практична цінність* полягає в розробці методичних вказівок для підвищення обізнаності персоналу у питаннях протидії методам соціального інжинірингу, а також у розробці опитувальника для аналізу рівня обізнаності.

Усі проведені операції зроблено виключно в рамках дослідження для визначення ступеню обізнаності користувачів. Усі конфіденційні дані користувачів, такі як паролі, не викладаються і не зберігаються у відкритому вигляді задля забезпечення їх захисту. Організатор дослідження залишає за собою право на зберігання, обробку та публікацію усіх зібраних даних, відповідно до умов користування сервісом, кожен з користувачів, що надсилав дані, попереднього погоджувався з умовами користування сервісом.

## 2. ТЕОРЕТИЧНІ ОСНОВИ ДОСЛІДЖЕННЯ

В контексті інформаційних технологій соціальний інжиніринг — це сукупність підходів прикладних соціальних наук, які орієнтовані на цілеспрямовану зміну організаційних структур, що визначають людську поведінку і забезпечують контроль за нею, або комплексний підхід до вивчення і зміни соціальної реальності, заснований на використанні інженерного підходу і наукомістких технологій.

Соціальний інжиніринг застосовується для:
– збору відомостей про мету підприємства;
– отримання конфіденційної інформації;
– прямого доступу до системи.

У сфері інформаційної безпеки термін «соціальний інжиніринг» (іноді — «соціальна інженерія») використовується для опису науки і мистецтва психологічної маніпуляції. За статистикою аналітичного центру компанії Infowatch, 55% збитків, пов'язаних з порушеннями інформаційної безпеки, виникають з вини співробітників, що підпали під вплив соціальних інженерів [8].

В освітній практиці ідеї соціального інжинірингу реалізуються шляхом застосування сучасних освітніх технологій і активних методів навчання, а також за допомогою «насичення» навчального процесу дисциплінами соціологічного і організаційного циклу, у тому числі:
– теорія і методи соціального інжинірингу;
– діагностика організацій;
– прогнозування і моделювання розвитку організацій;
– організаційне проектування і програмування;
– соціальне планування;
– впровадження соціальних нововведень в організації;
– практикум з соціальних технологій;
– методи вирішення конфліктів.

В основному інциденти соціального інжинірингу, пов'язані з діями персоналу, відбуваються через низький рівень обізнаності користувачів. Таким чином, навчаючи своїх співробітників основним правилам в області інформаційної безпеки, організації можуть значно знизити ризик порушення інформаційної безпеки. Не дарма, навчання персоналу — одна з основних вимог міжнародного стандарту управління інформаційною безпекою ISO/IEC 27001 [9].

Особливості атак з використанням людського фактору:
– не вимагають значних витрат;
– не вимагають спеціальних знань;
– можуть тривати протягом тривалого терміну;
– важко відслідковуються.

Людина часто набагато більш вразлива, ніж система. Саме тому соціальний інжиніринг спрямований на отримання інформації за допомогою людини, особливо в тих випадках, коли неможливо отримати доступ до системи (наприклад, комп'ютер з важливими даними відключений від мережі).

У соціальному інжинірингу є кілька технік, що використовуються для досягнення поставлених завдань. Всі вони засновані на помилках, що допускаються людиною в поведінці [10].

До технік соціального інжинірингу відносяться:

1. *Фішинг-атаки* — це найпопулярніший вид шахрайства в соціальному інжинірингу. Фішингова атака є незаконне отримання конфіденційних даних користувачів (логіна і пароля). Часто фішингові листи написані погано і містять граматичні помилки. У цих листах зловмисники вказують гіперпосилання на копію сайту (наприклад, поштового клієнта) з формою, де необхідно ввести свій логін, пароль і іншу особисту інформацію. Наприклад, фішинг застосовується для збору логінів і паролів користувачів шляхом розсилки листів і повідомлень, що спонукають жертву повідомити необхідну інформацію. Убезпечити себе від зловмисників можна ігноруванням листів від невідомих адресатів.

2. *Претекстинг* — атака, проведена за заздалегідь підготовленим сценарієм. Такі атаки спрямовані на розвиток почуття довіри жертви до зловмисника. Атаки зазвичай здійснюються по телефону. Цей метод часто не вимагає попередньої підготовки і пошуку даних про жертви. Претекстинг полягає у видачі жертви себе за іншу людину для отримання бажаних даних. Отримати інформацію про людину можна через джерела відкритого доступу, в основному зі сторінок соціальних мереж.

3. *Троянський кінь* використовує такі якості потенційної жертви, як цікавість і жадібність. Соціальний інженер відправляє електронного листа з безкоштовним відео або оновленням антивірусу у вкладенні. Жертва зберігає вкладені файли, які насправді є троянськими програмами. Така техніка залишиться ефективною до тих пір, поки користувачі продовжують бездумно зберігати або відкривати будь-які вкладення.

4. *Квіпрокво* (quid pro quo). При використанні цього виду атаки зловмисники обіцяють жертві вигоду в обмін на факти. Наприклад, зловмисник дзвонить в компанію, представляється співробітником технічної підтримки і пропонують встановити «необхідне» програмне забезпечення. Після того, як отримано згоду на установку програм, порушник отримує доступ до системи і до всіх даних, що зберігаються в ній.

5. *Зворотний зв'язок* має на увазі несанкціонований прохід зловмисника разом з законним користувачем через пропускний пункт. Такий спосіб не можна використовувати в компаніях, де співробітникам необхідно використовувати пропуски для входу на територію підприємства

6. Однією з технік соціального інжинірингу є *плечовий серфінг*. Він застосовується в транспорті, в кафе та інших громадських місцях, що дозволяють через плече жертви спостерігати за комп'ютерними пристроями і телефонами. Бувають ситуації, в яких користувач сам пропонує шахраєві необхідну інформацію, будучи впевненим у порядності людини. У такому випадку говорять про зворотній соціальний інжиніринг.

7. Загрози при використанні служби *миттєвого обміну повідомленнями*. Користувачі швидко оцінили зручність обміну повідомленнями в режимі реального часу за допомогою мереж Skype, Viber, WhatsApp, Telegam та ін. Доступність і швидкість такого способу спілкування робить його відкритим для всіляких атак. Для безпеки варто ігнорувати повідомлення від невідомих користувачів, не повідомляти їм особисту інформацію, не переходити за надісланими посиланнями.

Очевидно, що соціальний інжиніринг може завдати величезної шкоди будь-якій організації. Саме тому необхідно вживати всіх можливих заходів, щоб запобігти атак на людський фактор.

Спочатку завжди формується мета впливу на той чи інший об'єкт. Під «об'єктом» мається на увазі жертва, на яку націлена атака зловмисника.

Потім збирається інформація про об'єкт, з метою виявлення найбільш зручних мішеней впливу.

Після цього настає етап, який психологи називають атракцією. Атракція (від лат. attrahere 'залучати, притягати') — це створення необхідних умов для впливу зловмисника на об'єкт.

Примушення до потрібної для соціального хакера дії зазвичай досягається виконанням попередніх етапів, тобто після того, як досягнута атракція, жертва сама створює необхідні зловмиснику дії. Однак в ряді випадків цей етап набуває самостійну значимість, наприклад, тоді, коли примушення до дії виконується шляхом введення в транс, психологічного тиску і т. ін.

Всі атаки соціальних хакерів укладаються в одну досить просту схему (рис. 2).

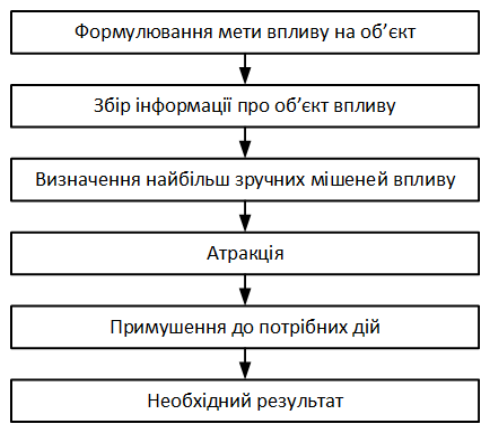

*Рис. 2. Основна схема впливу в соціального інжинірингу*

### 3. МЕТОДИКА ДОСЛІДЖЕННЯ

Зазвичай для зручності людей більшість публічних Wi-Fi мереж залишають відкритими, що робить їх гарним місцем для проведення різноманітних атак. Цей факт і надихнув на проведення даного дослідження.

Отже було вирішено створити відкриту Wi-Fi мережу для збирання даних з жертв наступним чином:

– можливість підключитися до бездротової мережі для будь-кого;

– псевдо-інтерфейс для реєстрації користувача в мережі, основним завданням якого буде збір даних про жертву включаючи дані які вона нам надасть та дані які ми отримаємо від браузера жертви, а саме User-Agent та Cookies для домену на який хотіла зайти жертва, що дозволить нам використовувати її автентифікацію на даному домені;

– налаштувати обладнання таким чином, щоби воно могло працювати в повністю автономному режимі;

– розмістити обладнання в місцях скупчення людей;

– забрати обладнання через тиждень (для досягнення автономності треба використовувати акумуляторні батареї великої ємності).

В експерименті задіяне наступне апаратне забезпечення:

– мініатюрний одноплатний енергоефективний комп'ютер на базі архітектури ARM з можливістю підключення пристроїв по інтерфейсу USB;

– кабель для подачі живлення типу USB-MicroUSB;

– портативний акумулятор (PowerBank) ємністю в 10000 мА/год. з USB-інтерфейсом;

– безпроводовий мережевий адаптер стандарту 802.11n з інтерфейсом USB та зовнішньою антеною;

– подовжувач USB-USB для зручності розташування елементів;

– карта пам'яті MicroSDHC Class 10 об'ємом 16 ГБ.

Для тестового стенду було вибрано наступне обладнання: Raspberry Pi 3 Model B, SanDisk MicroSDHC 16Gb Class 10, Trust PowerBank 10000 mAh, Tp-Link TL-WN722N v3. На рис. 3 зображено тестовий стенд у зібраному та увімкненому вигляді.

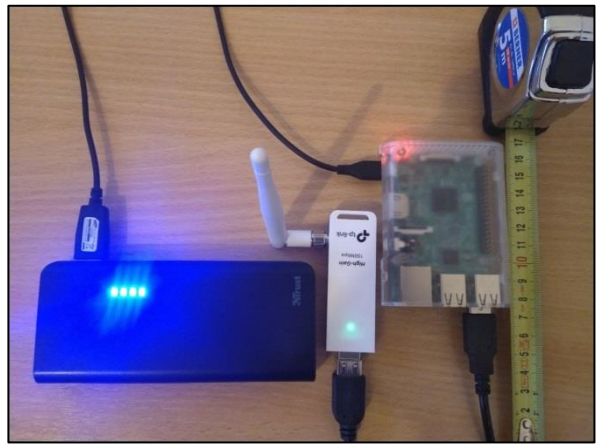

*Рис. 3. Зовнішній вигляд тестового стенду*

Для реалізації підробленої точки доступу та фішингового інтерфейсу на ній було визначено наступний інструментарій програмного забезпечення:

– hostapd — сервіс точки доступу Wi-Fi;

– dnsmasq — сервер DHCP та DNS;

– lighttpd — веб-сервер;

– PHP — мова програмування зі сторони веб-серверу;

– стек HTML, CSS та JavaScript для представлення в браузері;

– SQLite — база даних для зберігання даних.

В рамках проведення експерименту було вибрано три профілі вищих навчальних закладів (ВНЗ):

– технічний (Державний університет телекомунікацій, Київ);

– гуманітарний (Київський університет імені Бориса Грінченка, Київ);

– змішаний (Національний університет «Львівська політехніка», Львів).

Відповідно до кожного з них було розроблено окрему фішингову веб-сторінку (на рисунку 4 показаний приклад сторінки для Київського університету ім. Б. Грінченка).

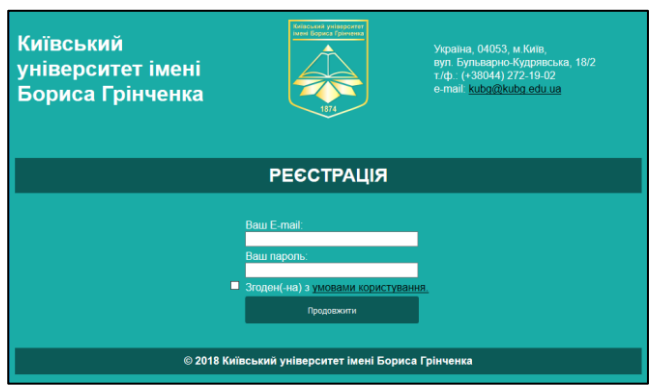

*Рис. 4. Приклад підробленої веб-сторінки для Київського університету ім. Б. Грінченка*

Для законності даного дослідження було розроблено умови користування сервісом (політику конфіденційності). Погодження з політикою конфіденційності сервісу користувачем перед відправленням даних надає організатору законні підстави на зберігання, обробку та публікацію цих даних. Точка доступу не мала доступу до інтернету, тому дані збиралися лише підчас першого підключення, дані зберігалися лише у вигляді статистики.

## 4. РЕЗУЛЬТАТИ ДОСЛІДЖЕННЯ

Під час аналізу результатів виникла необхідність в автоматизації процесу обробки інформації та знаходження зв'язків між даними. Зокрема для створення відповідностей між MAC та IP-адресами та діями на веб-сервері, було написано наступну програму на C#/.NET v. 4.5.

В дослідженні були отримані технічні дані (операційна система, версія браузеру, виробник мобільного пристрою тощо), дані поведінки (повторне підключення) і дані користувача (електронна пошта, паролі, кукі, запит до сайту-цілі). З усіх даних найбільшу цінність для дослідження соціального інжинірингу мають лише дані про поведінку і персональні дані, якими користувачі погодилися поділитися самі.

Якісним показником доступності інтернету є відсоток повторних підключень. З діаграми на рис. 5 видно, що кількість спроб повторного підключення не залежить від профілю вищого навчального закладу, а лише від доступності альтернативних безпроводових мереж.

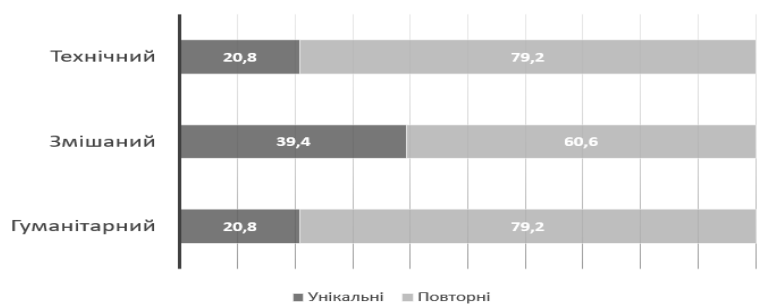

*Рис. 5. Статистика підключень*

Статистика легкості, з якою діляться користувачі своєю електронною адресою і навіть паролями, приведена на рис. 6. Тенденція показує збільшення відсотку наданих персональних даних у студентів гуманітарного профілю, але все одно довіра до невідомих відкритих мереж досить висока і серед студентів технічних ВНЗ. Високий відсоток введення пароля обумовлений введенням неіснуючих паролів.

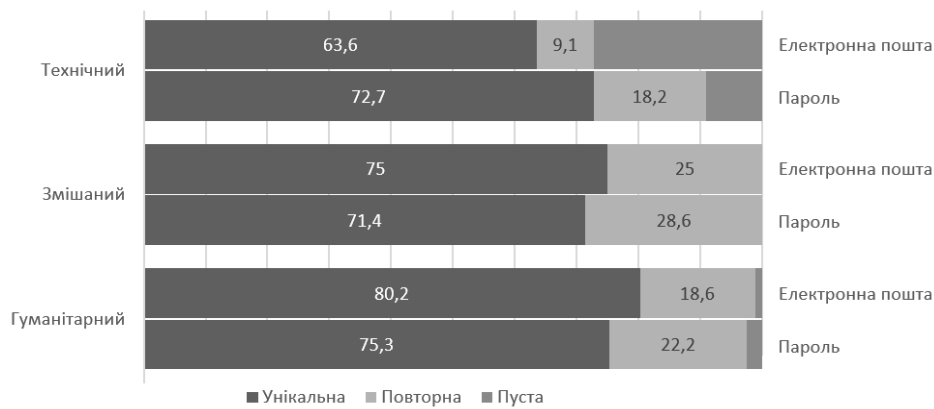

*Рис. 6. Статистика введення персональних даних (електронної пошти і пароля)*

Осторонь стоїть питання відкритості кукі, бо для отримання більшого функціоналу від веб-ресурсів більшість користувачів дозволяють обмінюватися цими даними без окремих запитів на передавання даних. З рис. 7 видно, що кількість користувачів, які відкривають доступ до кукі лише частково залежить від профілю ВНЗ. Тому питання відкритості кукі скоріш треба розглядати як загальну небезпеку обміну даними, а не тільки як аспект соціального інжинірингу.

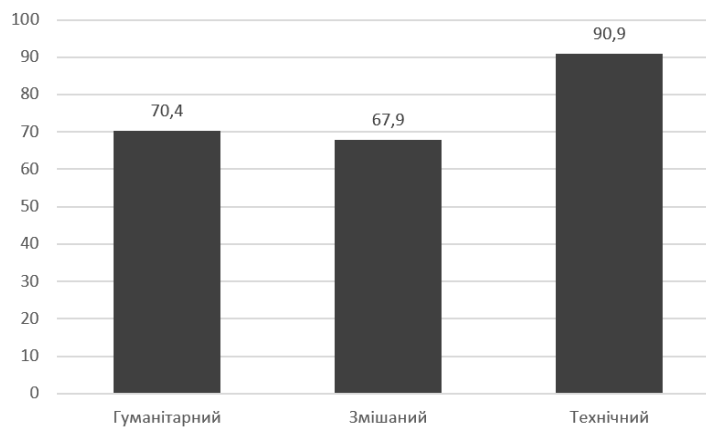

*Рис. 7. Статистки по дозволу роботи з кукі*

## 5. ВИСНОВКИ ТА ПЕРСПЕКТИВИ ПОДАЛЬШИХ ДОСЛІДЖЕНЬ

В ході проведеної роботи виявилися проблеми в конфігурації програмного забезпечення lighttpd. За замовчуванням, даний веб-сервер не використовує логування усіх звернень до нього, на відміну від Apache httpd. Тому такі статистичні дані, як виробники пристроїв, з пристроїв, що надавали дані веб-серверу, недоступні для місця проведення збору інформації в Державному університеті телекомунікацій. Даний факт був взятий до уваги і кількість місць збору інформації було збільшено до трьох.

Під час роботи було вирішено ряд проблем, пов'язаних із логуванням, автоматизацією аналізу та обробки даних, налаштуванням адресації IPv4 мережі, перехоплення усіх запитів від користувачів, та інші. Вирішені питання з узгодженням зв'язків між даними, зібраними різним програмним забезпеченням.

Основне завдання роботи — дослідження обізнаності користувачів щодо соціотехнічних атак було вирішено і отримано достатньо статистичних даних на протязі проведення дослідження. Так як експериментальний стенд виконаний на широкодоступних компонентах і у програмному середовищі з відкритим вихідним кодом, то такий вид атак можна легко відтворити будь-кому, що має середню технічну освіту за напрямками комп'ютерної інженерії і захисту інформації.

В результаті експериментів видно, що обізнаність користувачів навіть технічних спеціальностей недостатня, тому потрібно приділяти окрему увагу до розробки методик підвищення рівня обізнаності користувачів та зменшення кількості потенційних атак на об'єкти інформаційної діяльності.

### ПОДЯКА

### СПИСОК ВИКОРИСТАНИХ ДЖЕРЕЛ

UDC 316.776:004.58

**Volodymyr Yu. Sokolov**
MSc, senior lecturer
Borys Grinchenko Kyiv University, Kyiv, Ukraine
OrcID: 0000-0002-9349-7946
*vladimir.y.sokolov@gmail.com*

**Davyd M. Kurbanmuradov**
MSc
State University of Telecommunications, Kyiv, Ukraine
OrcID: 0000-0003-4503-3773
*rockyou@protonmail.com*

# METHOD OF COUNTERACTION IN SOCIAL ENGINEERING ON INFORMATION ACTIVITY OBJECTIVES

**Abstract.** The article presents a study using attacks such as a fake access point and a phishing page. The previous publications on social engineering have been reviewed, statistics of break-ups are analyzed and directions and mechanism of realization of attacks having elements of social engineering are analyzed. The data from the research in three different places were collected and analyzed and the content statistics were provided. For comparison, three categories of higher education institutions were chosen: technical, humanitarian and mixed profiles. Since the research was conducted in educational institutions during the week, most students in the experiment and graduate students took part in the experiment. For each educational institution, a registration form template was created that mimicked the design of the main pages. Examples of hardware and software implementation of a typical stand for attack, data collection and analysis are given. In order to construct a test stand, widely available components were chosen to show how easy it is to carry out attacks of this kind without significant initial costs and special skills. The article provides statistics on the number of connections, permission to use the address of the e-mail and password, as well as permission to automatically transfer service data to the browser (cookies). The statistics are processed using specially written algorithms. The proposed approaches to solving the problem of socio-technical attacks can be used and implemented for operation on any objects of information activity. As a result of the experiments, it is clear that the awareness of users of even technical specialties is not enough, so one needs to pay particular attention to the development of methods for raising awareness of users and reducing the number of potential attacks on objects of information activity.

**Keywords:** social engineering; attack; fishing; access point; protection of personal information.